# Graphenes flakes under controlled biaxial deformation

Charalampos Androulidakis[1,2], Emmanuel N. Koukaras[1], John Parthenios[1], George Kalosakas[1,2,3], Konstantinos Papagelis[1,2], and Costas Galiotis[*1,4]

[1]*Institute of Chemical Engineering Sciences, Foundation of Research and Technology-Hellas (FORTH/ICE-HT), Stadiou Street, Platani, Patras, 26504 Greece*
[2]*Department of Materials Science, University of Patras, Patras, 26504 Greece*
[3]*Crete Center for Quantum Complexity and Nanotechnology (CCQCN), Physics Department, University of Crete, 71003 Heraklion, Greece*
[4]*Department of Chemical Engineering, University of Patras, Patras 26504 Greece*

[*]Corresponding author: c.galiotis@iceht.forth.gr or galiotis@chemeng.upatras.gr

## ABSTRACT

Thin membranes, such as monolayer graphene of monoatomic thickness, are bound to exhibit lateral buckling under uniaxial tensile loading that impairs its mechanical behaviour. In this work, we have developed an experimental device to subject 2D materials to controlled equibiaxial strain on supported beams that can be flexed up or down to subject the material to either compression or tension, respectively. Using strain gauges in tandem with Raman spectroscopy measurements, we monitor the G and 2D phonon properties of graphene under biaxial strain and thus extract important information about the uptake of stress under these conditions. The experimental shift over strain for the G and 2D Raman peaks were found to be in the range of $62.3 \pm 5$ cm$^{-1}$/%, and $148.2 \pm 6$ cm$^{-1}$/%, respectively, for monolayer but also bilayer graphenes. The corresponding Grüneisen parameters for the G and 2D peaks were found to be between $1.97 \pm 0.15$ and $2.86 \pm 0.12$, respectively. These values agree reasonable well with those obtained from small-strain bubble-type experiments. The results presented are also backed up by classical and *ab initio* molecular dynamics simulations and excellent agreement of $\Gamma$-E$_{2g}$ shifts with strains and the Grüneisen parameter was observed.

# INTRODUCTION

As is now well established, graphene is the first ever 2-dimensional crystal and is constituted of carbon atoms ordered in a honeycomb hexagonal lattice. The measured or predicted extraordinary properties of graphene, such as, its high carrier mobility, its high thermal conductivity, its high stiffness and strength among other things can or have already found useful applications in the field of electronics and sport [1-9]. Furthermore, as it has been postulated in a number of earlier publications, the application of strain alters the graphene lattice and can induce changes of the electronic properties of the material, while for multilayer graphene a band gap can be opened [10-14]. Finally, the use of graphene as a reinforcing agent in composite materials is also a very promising application field which is still at its infancy due to difficulties of handling and processing relatively large graphene membranes as reinforcements for suitable matrices.

The effect of uniaxial strain on single layer graphene has been studied experimentally by the imposition of axial loads upon supported graphene flakes on plastic bars which can be flexed up or down to subject graphene to compression and tension, respectively, for strain levels of up to ~1.5% [2-5,10]. In all these studies Raman spectroscopy has been employed to probe the shift of the Raman peaks with respect to the applied strain so as to monitor the phonon behaviour under mechanical stress and calculate the relative Grüneisen parameters of the material. Lee et al.[1] subjected a suspended graphene flake to biaxial tension by bending the flake by an AFM indenter. By considering graphene as a clamped circular membrane made by an isotropic material of zero bending stiffness, they converted by means of pertinent modelling the bending force vs. deflection curve to an "axial" stress-strain curve. This way they managed to confirm the extreme stiffness of graphene of 1 TPa and provided an indication of the breaking strength of graphene of 42 N m$^{-1}$ (or 130 GPa considering graphene thickness as 0.335 nm). Other true biaxial experiments were attempted by Zabel et al.[15] who employed graphene bubbles formed during the deposition of large graphene flakes on a oxidized silicon substrate, and having an estimated strain of ~ 1%, to study graphene under biaxial (e.g., isotropic) strain and by Metzger et al.[16] who studied adhered graphene to shallow depressions and finally Ding et al.[17] who employed a piezoelectric substrate to induce biaxial strains. Further details of these first attempts to induce biaxial strain fields to graphene are given later.

As is evident, biaxial deformation (stretching) is particularly relevant for thin films or membranes at all scales. In this context, CVD-grown graphenes that can be produced in macroscopic dimensions are quite important for numerous applications (e.g. electroactive screens, filters, coatings etc.) for which biaxial loading is required. To date the only cited work in this area is by Jie *et al.*[18] who employed a piezoelectric substrate to subject graphene to a small biaxial strain of – 0.071%. Interestingly, after the initial application of strain the deformation kept on increasing (creep phenomena) and this was attributed to internal movement at the grain boundaries. All the above techniques applied to exfoliated or CVD graphene, although diverse in conception and design, are characterized either by a total absence or limited ability to control and/or assess the applied strain levels, which in most cases were extremely low.

In this work a new experimental technique has been developed for subjecting 2-dimensional crystals (such as graphene, $MoS_2$, etc.) to controllable equibiaxial tensile strain gradients. The principle of this technique is based on the extension along two dimensions of the three-point bending configuration of plastic bars that have been employed for uniaxial loading as mentioned earlier. The added advantage of this approach is the fact that due to the symmetry of the loading procedure the term referring to the shear deformation potential (SDP) becomes zero and therefore the Poisson's ratio of the underlying substrate has no effect on the measured strain[4]. To accomplish this, a plastic substrate with cruciform shape, as shown in **Figure 1**, is symmetrically deformed about its centre, thus inducing an equi-biaxial strain at that position. The advantages of this technique compared to those attempted earlier are the following: (a) the biaxial strain is applied at a stepwise and controllable manner, (b) the strain can be directly measured using strain gauges, (c) the setup is capable of loading any 2-dimensional material at moderate strain levels and, finally, (d) the jig is housed under a Raman microscope that allows mapping of stress or strain with submicron resolution. Effectively, this means that the present method carries all the advantages and simplicity of the corresponding uniaxial technique, which has been successfully applied in a number of uniaxial strain studies on nanoscale materials such as graphene [2-5]. Furthermore, by employing electrical resistance strain gauges and by monitoring the shifts of the 2D and G Raman lines, information on the stress transfer efficiency upon application of equi-biaxial strain can easily be obtained, for simply supported as well as fully embedded graphene in polymer matrices.

In this work, we have examined several graphene samples within a range of thicknesses from monolayer up to nanographite, and have monitored in detail the response of the Raman phonons. To our knowledge this is the first time that phonon shifts for trilayer graphene are reported under biaxial deformation and this complements the prior work reported under uniaxial strain [19]. The effect of equibiaxial strain on the doubly degenerate Raman active $E_{2g}$ phonon at the Γ point has also been examined theoretically. From the phonon shifts the experimental Grüneisen parameters for the G- and 2D peaks can be retrieved. Finally, a method that utilizes atomic trajectories and velocities from classical and *ab initio* molecular dynamics (AIMD) simulations has been used for the implicit calculation of the Γ-$E_{2g}$ phonon frequency of graphene at finite temperature, thus providing a more realistic correspondence to experiments. The numerical results are in good agreement with the experimental measurements.

**RESULTS AND DISCUSSION**

In **Figure 1a** we present a sketch of our purpose-built experimental device and an actual photograph of its operation is shown in **Figure SI-3**. The substrate is a plastic cruciform that is pinned at its four edges. An adjustable screw under its geometric centre deflects the substrate upwards. This allows for an equi-biaxial tensile strain gradient to develop on the top surface of the plastic bar. The strain level at the geometrical centre is given by the following equation

$$\varepsilon_{xx} = \varepsilon_{yy} = \frac{3h\delta}{L^2} \tag{1}$$

Where $h$ is the thickness of the plastic bar (substrate–cruciform), $\delta$ is the deflection of the bar at its centre, and $L$ is the length between the two opposing pin edges, as shown in **Figure 1**.

As argued in earlier publications [4,6] the phonon modes of graphitic materials such as graphene and carbon fibres, are linearly related to applied stress or moderate strains. For the $E_{2g}$ mode the solution of the secular equation under an externally applied strain field yields[4]:

$$\Delta\omega_G^\pm = \Delta\omega_G^h \pm \Delta\omega_G^s = -\omega_G^0 \gamma_G (\varepsilon_{ll} + \varepsilon_{tt}) \pm \frac{1}{2}\omega_G^0 \beta_G (\varepsilon_{ll} - \varepsilon_{tt}) \tag{2}$$

where $\Delta\omega_G^h$ and $\Delta\omega_G^s$ are the shifts corresponding to the hydrostatic and shear (mode splitting) components of the strain and $\gamma_G$ is the Grüneisen parameter of the in-plane Raman active E$_{2g}$ phonon and $\beta_G$ is the shear deformation potential. As is well established, the Grüneisen parameter provides important information on the thermomechanical response of phonon modes. For our case here, there are certain advantages in calculating the Grüneisen parameter of the *G* peak of graphene by applying equibiaxial strain [4]. This is because under these conditions ($\varepsilon_{ll}=\varepsilon_{tt}$) the second term of equation (2) diminishes and the normalized shift is related to strain by:

$$\frac{\Delta\omega_{G;2D}}{\omega^0_{G;2D}} = -2\gamma_{G;2D}\varepsilon \qquad (3)$$

Thus for biaxial strain the shift of the Raman E$_{2g}$ frequency is related to applied strain through the corresponding Grüneisen parameter. Furthermore, no splitting of the peak due to lifting of degeneracy is induced and therefore the Poisson's ratio of the substrate is not required as in the case of uniaxial experiments [3-5]. The same relation applies for the 2D peak (see above equation)[4,15].

Metzger *et al.*[16] covered mechanically exfoliated monolayer graphene flakes over shallow depressions that were patterned on a SiO$_2$/Si wafer. The depressions had a square shape with 6 μm sides and 20 nm depth. The reported equi-biaxial strain was 0.066%. The Raman shifts and Grüneisen parameters (which they calculated using **eqn. 3**) for the G and 2D peaks are given in **Table I**. Zabel*et al.*[15] deposited mechanically exfoliated monolayer graphene flakes over a SiO$_x$/Si substrate. Having optically identified spherical graphene bubbles with diameters in the range of 5–10 μm, they reported average biaxial strains of ~ 1%. Using **eqn.3** the calculated Grüneisen parameters were smaller than those obtained by Metzger *et al.* but in good agreement with the experimental and theoretically values from uniaxial experiments given by Mohiuddin *et al*[4]. Perhaps the only work to date that demonstrates control (and with a very fine step) over biaxial strain applied on monolayer graphene flakes is by Ding *et al.*[17] who used for that purpose a piezoelectric actuator as the substrate. Strain levels in the range –0.15% to 0.1% were calculated (not measured) through **eqn.3** by measuring the peak shift and then converting to strain using the Grüneisen parameters estimated by Mohiuddin *et al.* for uniaxial tension (see **Table I**). Metten *et al.*[20] deposited monolayer graphene over circular pits with diameter of ~4 μm and using blister tests they obtained the corresponding Grüneisen parameters and calculated the Young's modulus. The response under high pressures of mechanically exfoliated graphene [21] and graphene grown from chemical vapor deposition [22] has also been studied employing a diamond anvil cell. Finally, theoretical calculations [4,23-26] based on density functional theory within the framework of the local density approximation (LDA) and generalized gradient approximation (GGA) are all in very good agreement and, as shown in **Table I,** produce similar values for the G-peak Grüneisen parameter.

In the experiments presented here, two types of glassy polymers were used as the substrate, PMMA (Poly-methyl methacrylate) and polycarbonate (PC). The PMMA bars can reach a biaxial strain level prior to failure of $\varepsilon_{xx} = \varepsilon_{yy} \sim 0.5\%$. To achieve larger strain levels one can switch to polycarbonate which has similar mechanical properties to PMMA but exhibits much higher stress and strain to failure. A strain gauge rosette was placed on the centre of the substrate in order to calibrate the developed strain on the beam surface. Several measurements were taken with strain gauges placed on several substrates on both PMMA and polycarbonate bars (**see image in Figure SI-3**). The measurements show that by bending the substrate upwards (at its geometric centre) the strains in the $x$ ($\varepsilon_{xx}$) and $y$ ($\varepsilon_{yy}$) directions are tensile and equal and this confirms the validity of the design considerations of our apparatus. Certainly, in future experiments the design itself can be easily modified by allowing for opposing pins to be adjustable. This would ultimately allow the application of either equi-biaxial, or non-equal biaxial strain, or even for subjecting the specimen to complex biaxial strain fields (eg. tensile in one direction and compressive in the other). This versatility may also prove very useful for the study in a controllable manner of other 2-dimensional highly anisotropic materials, such as $MoS_2$, black phosphorus, etc.

Several experiments were performed on simply-supported monolayer graphene flakes. In some cases a thin layer of SU-8 was spin coated over the polymer substrate, prior to the deposition of the graphenes, in order to improve the contrast and thus the optical visibility of graphene. The combinations examined were graphene resting on (a) PMMA, (b) PC, (c) PMMA/SU-8, and (d) PC/SU-8. The chosen polymers possess similar mechanical properties and, as the results clearly show, graphene adheres well to all of them. The results for all cases are presented in **Table II**. Although PC has the advantage of higher flexibility, however it exhibits a broad Raman peak in the vicinity of graphene G peak. This overlap makes the deconvolution of the two peaks very difficult and therefore in this case we only present the 2D peak shifts.

Prior to the experimental procedure, several Raman measurements were taken at central regions of the graphene flakes. For any given flake the strain distributions were found to be uniform with very small deviations of the Raman peaks. For each case only a negligible or small Raman wavenumber downshift was identified indicating that some of the flakes initially had a small residual tensile strain which apparently only affects the position of the zero strain developed in graphene and not the relationship between Raman frequency and strain in this situation[5,27].

In **Figure 2** results for the position of the 2D and G peaks versus the applied strain are presented (further details are given in the SI) together with their corresponding spectra at selected strain levels. The mean values of the shift of the phonon peaks per percent of strain were found to be –148.2±6 cm$^{-1}$/% and –62.3±5 cm$^{-1}$/% for the 2D and G peaks respectively. The corresponding Grüneisen parameters were estimated to be 2.86±0.12 and 1.97±0.15 in reasonable agreement with the reported experimental and theoretical values (see **Table I**). We note here that although the shift of the 2D peak of the present experiments is comparable to the shifts reported by Zabel *et al.*[15] for the graphene balloons, however the Grüneisen parameter of the present study is larger.

The maximum strain achieved without any significant graphene failure was ~0.42%, at which point the PMMA cruciform broke. This strain relaxation may indicate local slippage or interface failure [28] due to the fact that only van der Waals forces are activated for the transfer of stress or strain from the cruciform to the graphene itself. The maximum strain achieved with no sign of failure, as identified by the linearity of the Pos(2D or G)–strain curves, was ~0.28% for monolayer (**Figure 2** and **Figure SI-1c**) and 0.42% for bilayer (**Figure 3**) graphene. For the embedded bilayer graphene no indications of failure was observed as in this case the relation between the Pos(2D or G) and the applied strain was found to be linear up to specimen fracture.

In the present results no splitting was observed for the 2D or the G peak. This indicates that there is no effect of the substrate's Poisson ratio on the strain sensitivity. In most cases a small increase in the FWHM was observed with increasing strain (**figures SI-2**). Despite this increase the lack of any observed splitting reflects the conservation of the E$_{2g}$ phonon symmetry which is manifested here by the equal shift rates of the G$^+$ and G$^-$ sub-peaks, as would be expected when the applied strain is equi-biaxial.

We now turn our attention to bilayer graphene. We note that the only available results for comparison with the literature on shift rates for bilayer graphene are for bilayer graphene bubbles and balloons from Ref. 15, but such high shift rates similar to monolayer graphene are for the first time reported herein. In this case in order to achieve efficient stress transfer to the bilayer graphene the graphene specimens were fully embedded into the polymer matrix. In **Figure 3** the position of the G and 2D Raman peaks versus the applied strain are plotted. The measured 2D peaks of the spectra are characteristic of Bernal (AB) stacking [7] and the 2D peak is fitted with four Lorentzian curves. The shifts rates of the four 2D components are –149.3, –152.6, –162.6, –152.6 cm$^{-1}$/% for the $D_{22}$, $D_{21}$, $D_{12}$, $D_{11}$ sub-peaks, respectively. The evolution of the 2D Raman peak is also presented for various strain levels. The consistent form of the 2D spectra indicates that the AB stacking is preserved throughout the range of applied strains due to the equi-biaxial strain field. This means that there is no relative slipping between the individual layers that form the bilayer and both are stressed equally as a result from the direct attachment to the polymer. Thus, the strain level achieved here is small for inducing these kind of structural changes as was pointed out and elsewhere [15]. This has to be contrasted with uniaxial tensile experiments for which loss of Bernal stacking was noted at strain levels as low as 0.4%[29]. The G peak exhibits a shift rate of –57.2 cm$^{-1}$/% which is similar to values reported for monolayer graphene by us and others[15,17]. Overall the consistency of the results obtained for both monolayer and multi-layer graphenes demonstrates clearly the versatility of the proposed biaxial jig and its suitability for any 2D material.

Thicker graphene flakes were additionally examined that include trilayer (3LG) graphene, few layer graphene (FLG) and nanographite (NG). To our knowledge, results from biaxially strain controlled experiments have not been previously reported for graphenes of these thicknesses. The corresponding 2D spectra are presented in **Figure 4**. The trilayer flake that we examined is of ABA stacking.

For trilayer and few-layer graphene a change in slope is noted at ~0.15%, and for nanographite at ~0.23%. The observed changes in the slopes indicate that in these systems under biaxial strain the stress transfer efficiency is affected by the strain level. Since in effect no changes in the graphene/polymer adhesion are expected by the results presented for the monolayer and bilayer graphenes that exhibit identical interfaces, it can be assumed that the change of slope indicates cohesive failure or slipping in these systems. This is also corroborated by the fact that both G and 2D phonons exhibit the same behaviour under strain and therefore the effect cannot be due to phonon anomalies but to problems related to the specimen itself. The monolayer and bilayer graphenes do not suffer from cohesive failure or slippage, in contrast to the thicker graphenes. This is a result from the directly stress transfer from the polymer to the graphene layers in these cases, in contrast to the few layer flakes where the inner layers are stressed only by the stress transfer of the weak van der Waals forces that bond the mono-layers. In **Table III** we provide the values of the slopes near the origin, i.e. that correspond to strain ranges prior to the onset of the non-linear behaviour. In **Figure 5** we show the shift rates for the trilayer, few layer and nanographite specimens. The shift rates of both the 2D and G peaks decrease in value as the thickness increases (see **Table III**). For comparison purposes the values are given in **Table III**. Regarding the Grüneisen parameter the 2D peak of bilayer and multilayer graphenes consists of multiple components that exhibit different shift behaviour[30]. Thus the definition of a single Grüneisen parameter in these cases is not possible and therefore any comparison with the values obtained from the monolayer graphene is problematic. Nevertheless, even in these cases one can estimate an averaged shift rate by fitting all of the components to a single Lorentzian and this gives rise to an "average" Grüneisen parameter. Currently a systematic work is under way to identify the origin of the stress transfer efficiency in these systems.

In order to gain an insight in the effect of biaxial strain upon the $E_{2g}$ phonon frequency, we performed classical and *ab initio* molecular dynamics (MD) simulations. From the MD simulations we obtained the frequencies of the $E_{2g}$ phonon by suitable processing of the atomic trajectories and velocities. This approach has the advantages of being applicable at finite temperatures as well as accounting for anharmonic effects. To the best of our knowledge such results have not been reported in the literature. A detailed description of the method is given in Ref. [31].

For the classical MD simulations we employed the AIREBO [32], Tersoff-2010 [33](a reparameterisation of the original Tersoff[34,35] potential by Lindsay and Broido) and LCBOP [36] potentials. The original Tersoff potential was not considered since it unrealistically overestimates the frequencies of the LO/TO dispersion curve branches (see Ref. [31]). Details on the setup of the calculations are given in the Methods section. In **Figure 6** we have plotted the $\Gamma$-$E_{2g}$ frequencies that correspond to a temperature of $T = 300$ K for strains up to 2%. All of the potentials produce the expected decreasing linear dependence. We can see that the Tersoff-2010 and LCBOP potentials produce slopes of –60.7 cm$^{-1}$/% and –59.7 cm$^{-1}$/%, respectively, which are in very good agreement to the experimental values. The AIREBO potential overestimates the shift rates with a slope value of –78.4 cm$^{-1}$/%. This performance complements the trends noted in the overall performance of these potentials on describing the dispersion curves of graphene at $T$ = 300 K, as shown in Ref. [31], specifically that LCBOP produces the most accurate dispersion curves followed by Tersoff-2010. This also holds for the optical branches which are of interest here. In comparison the AIREBO potential significantly overestimates the highest optical branches.

The *ab initio* MD simulations were performed with the electronic structure computed using density functional theory (DFT) within the local density approximation (LDA), at a temperature of $T$ = 300 K. Details on the calculations are given in the Methods section. The shift rate of the $\Gamma$-$E_{2g}$ mode on strain is computed at –60.5 cm$^{-1}$/%, and the corresponding Grüneisen parameter is 1.82, which are in the experimentally expected range reported here and by other works (**Table I**) and also in very good agreement with the theoretical works shown in **Table I**. In comparison, the results for both the shift rate and the Grüneisen parameter using the Tersoff-2010 potential are in excellent agreement, followed closely by those using the LCBOP potential.

As mentioned earlier, few results have been reported on biaxial deformation of graphene which may be attributed to the difficulties and complexity in designing and performing such experiments. Most data reported so far correspond to the imposition of biaxial strain either by bending or by the formation of a "bubble" and the calculation in some cases of the phonon shifts for the 2D and G peaks. It is, however, worth noting here that for the phonon shifts a comparison with uniaxial experiments is rather difficult to be made because the strain levels reported for the pure biaxial experiments are very low. Thus one way of assessing the various reported data obtained from either freely-supported or suspended graphene is via the measured (or extracted) Grüneisen parameters for the G and 2D peaks. Indeed, since the Grüneisen parameter is a universal constant for the phonon characteristics of a perfect 2D crystal (such as graphene) any comparison of the actual values will highlight the problems encountered in this field vis-à-vis the methodology proposed here.

Using the method presented above, the shift rates of the G and 2D peaks, as well as the Grüneisen parameters $\gamma_G$ and $\gamma_{2D}$ were determined from the experimental results and compared with theoretical predictions. Very good agreement was found between experiment and theory, as well as with the results of the other studies for similar strain ranges.

In **Figure 7** the Grüneisen parameters for the G versus the 2D phonon as obtained experimentally in this work and also those reported in the literature are plotted. For the uniaxial experiment reported in Ref. 6 we have measured (see SI) the Poisson's ratio of the PMMA/SU-8 system which was found ~0.35. Since this value is slightly higher than that used in Ref. 6 we re-calculated the Grüneisen parameter for the 2D phonon and a value of 3.65 is obtained. The values for $\gamma_{2D}$ that we present in **Figure 7**, show a considerable scatter, as opposed to $\gamma_G$, in which case any scatter seems to emanate mainly from experimental error. It seems overall that there is certain discrepancy between $\gamma_{2D}$ values deduced from uniaxial experiments with supported samples vis-à-vis values obtained from biaxial experiments with suspended samples. This may be another indication of the effect of underlying substrate on the 2D peak and the Dirac points' position which has an effect on the derived Grüneisen parameter[4,37]. This is currently under investigation in an attempt to understand fully the differences observed for the $\gamma_{2D}$ values.

**CONCLUSIONS**

In the present study a new experimental technique is introduced that employs a device designed for subjecting any 2D crystal to controlled biaxial tensile deformations. The device can be easily handled and possesses all the features of the corresponding uniaxial devices. Graphene flakes of various thicknesses, ranging from monolayer to nanographite, were examined under biaxial strain in tandem with Raman spectroscopy measurements. For the monolayer graphene the 2D and G band shift rates were found to be $-148.2 \pm 6$ and $-62.3 \pm 5$ cm$^{-1}$/%, respectively. These rates correspond to Grüneisen parameters of $\gamma_{2D} \sim 2.86 \pm 0.12$ and $\gamma_G \sim 1.97 \pm 0.15$ which represent the best estimates to date of the Grüneisen constants since they have been derived from direct mechanical measurements at a whole range of strain values. Experiments were also conducted on multilayer graphenes; shift rates of the 2D and G Raman peaks for the bilayer were similar to the monolayer with values of –154.3 and –57.2 cm$^{-1}$/%, respectively, whereas for few layer graphenes (from trilayer to nanographites) a reduction of above rates were observed due possibly to cohesive failure within the flakes. These results are in good agreement with both classical and *ab initio* theoretical results at finite temperature obtained by a method based on molecular dynamics simulations. The $\Gamma$-$E_{2g}$ mode Grüneisen parameter and shift rate from the *ab initio* calculations are 1.8 and –60.5 cm$^{-1}$/%, respectively.

**METHODS**

**Sample preparation.** Cruciform polymer substrates, as schematically shown in **Figure 1**, were prepared by cutting rectangular polymer sheets. The substrates had final dimensions with a length of 13 cm and width of 1.1 cm. In some cases the substrates were covered on the top by a 200 nm thick layer of SU8 photoresist (SU8 2000.5, MicroChem). The embedded flakes were prepared by spin coating an additional PMMA layer on the top (with a thickness ~150 nm), having previously identified the graphene flake to be tested. The graphene flakes were first located using an optical microscope and the number of layers was identified with Raman measurements. No D peak was observed in the Raman spectra indicating the high quality of the exfoliated graphenes.

The monolayer flakes were subjected to biaxial strain using the two dimension three-point-bending jig as discussed earlier. The strain was applied incrementally with a step of 0.025% for PC and of 0.028% for PMMA. This small difference is the result of the PMMA polymer that was used being slightly thicker than the PC, and the jig calibration step. At every loading step measurements for the Raman 2D and G peaks were carried out. A laser of $\lambda_{laser}$=785 nm (1.58 eV) excitation was used to obtain the Raman spectra. The Raman measurements were taken at the centre of the flakes to avoid edge effects [38].

**Classical Molecular Dynamics.** The MD simulations were performed at 300 K using a triclinic computational cell of $20 \times 20$ unit cells (overall 800 carbon atoms) and periodic boundary conditions. The procedure followed for the simulation is as follows. The computational cell was initially relaxed for each potential (AIREBO, Tersoff-2010, and LCBOP). Randomized velocities (for all three dimensions) were attributed to the atoms, within a Gaussian distribution, corresponding to a temperature of $T$ = 300 K and initial equilibration at constant energy (microcanonical, NVE ensemble) was performed. The corresponding lattice parameter (for each potential) was computed by performing a subsequent equilibration within the isothermal–isobaric ensemble (NPT) at zero pressure. The lattice constant was taken as the average over these steps. A computational cell was created anew using the corresponding constant lattice for each potential. This cell was used in the NVE simulations from which the trajectories and velocities were acquired for the zero strain level. Transition to the next strain levels was performed by deforming the unit cell with care on uniformity, specifically, the unit cell edges were equally elongated and the angle was maintained at 60° by elongating the tilt of the triclinic cell to half of that of the $a_1$ axis (along the *x*-axis). After each elongation thermal equilibration was performed prior to the main NVE simulations. The time step was 0.05 fs and the NVE simulations were run ~32ps (for recording the trajectories and velocities used in the follow up *k*VACS (*k*-space velocity autocorrelation function) method to extract the phonon frequencies). To produce more reliable statistics 10 realizations were performed at each strain level from which an average velocity autocorrelation sequence was computed. All of the classical molecular dynamics simulations were performed using LAMPPS [39].

***Ab Initio* Molecular Dynamics.** The AIMD simulations were performed in the Born-Oppenheimer approximation using the QUICKSTEP method [40] that is based on the Gaussian and plane-wave approach (GPW)[41]. With this method two basis sets are employed; a localized Gaussian basis set for the Kohn-Sham orbitals (for the wavefunction) and planewaves for the electron density. The calculations were performed using the double-ζvalence plus polarization (DZVP) basis set with LDA-optimized Goedecker-Teter-Hutter (GTH) norm-conserving pseudo-potentials[42,43]. The planewave cutoff was set to 500 R y and four valence electrons were used for the carbon atoms ($2s^2 2p^2$). A triclinic computational cell of 6×6 unit cells (overall 72 carbon atoms) and periodic boundary conditions were used in the *x*–*y* plane. A vacuum of 25 Å between periodic images of the graphene is considered in the *z*-direction. The procedure followed was analogous to that for the classical molecular dynamics simulations. The time step was 0.8 fs and the NVE simulations were run ~20ps (for recording the trajectories and velocities used in the follow up *k*VACS method to extract the phonon frequencies). All of the *ab initio* molecular dynamics simulations were performed using CP2K [40].

**Acknowledgments**

This research has been co-financed by the European Union (European Social Fund-ESF)and Greek national funds through the Operational Program ''Education and LifelongLearning'' of the National Strategic Reference Framework (NSRF) - Research Funding Program: ERC-10

''Deformation, Yield and Failure of Graphene and Graphene-based Nanocomposites''. We acknowledge computation time provided by the Cy-Tera facility of the Cyprus Institute under the project CyTera. Usage of the Metropolis cluster at the CCQCN of the University of Crete is also acknowledged.Mr. C. Tragoulias is thanked for manufacturing the biaxial strain device. G.K. acknowledges European Union's Seventh Framework Programme (FP7-REGPOT-2012-2013-1), grant agreement no. 316165.Finally, the authors acknowledge the financial support of the Graphene FET Flagship (''Graphene-Based Revolutions in ICT And Beyond''- Grant agreement no: 604391) and of the European Research Council (ERC Advanced Grant 2013) via project no. 321124, "Tailor Graphene".

**Author Contributions**

C.G. supervised and conceived the project. Ch. A. designed the experimental setup, prepared the samples and performed the experiments. Ch. A., J. P., C. G. analyzed the data and interpreted the results. E.N.K., G.K., K.P. derived the computational modelling and E.N.K. performed the runs. Ch. A., E. N. K. and C. G. wrote the paper. All authors reviewed and commented on manuscript.

**Competing financial interests:** The authors declare no competing financial interests.

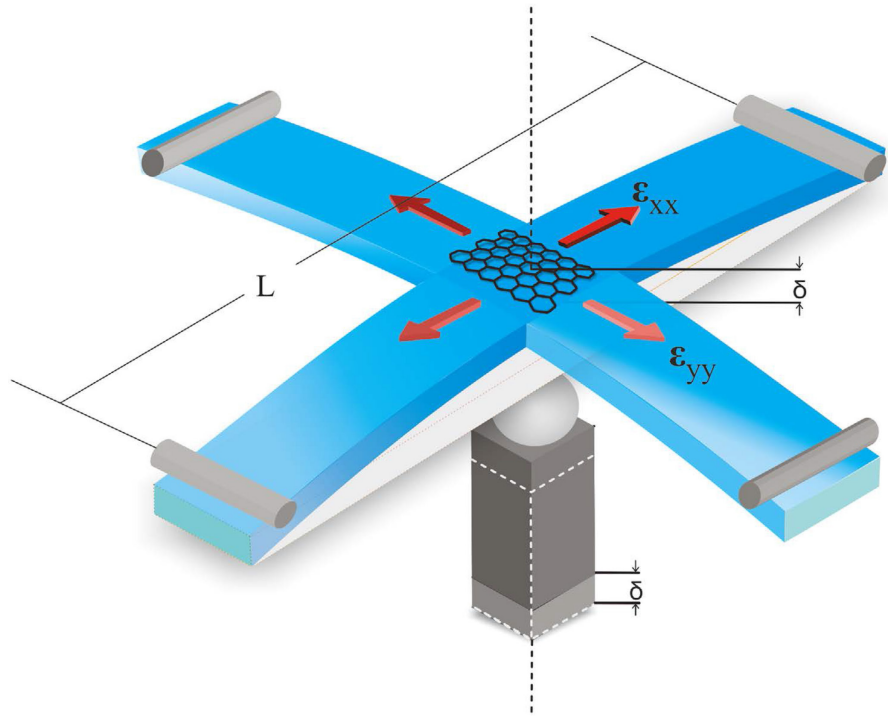

**Figure 1.** Schematic of the biaxial strain apparatus.

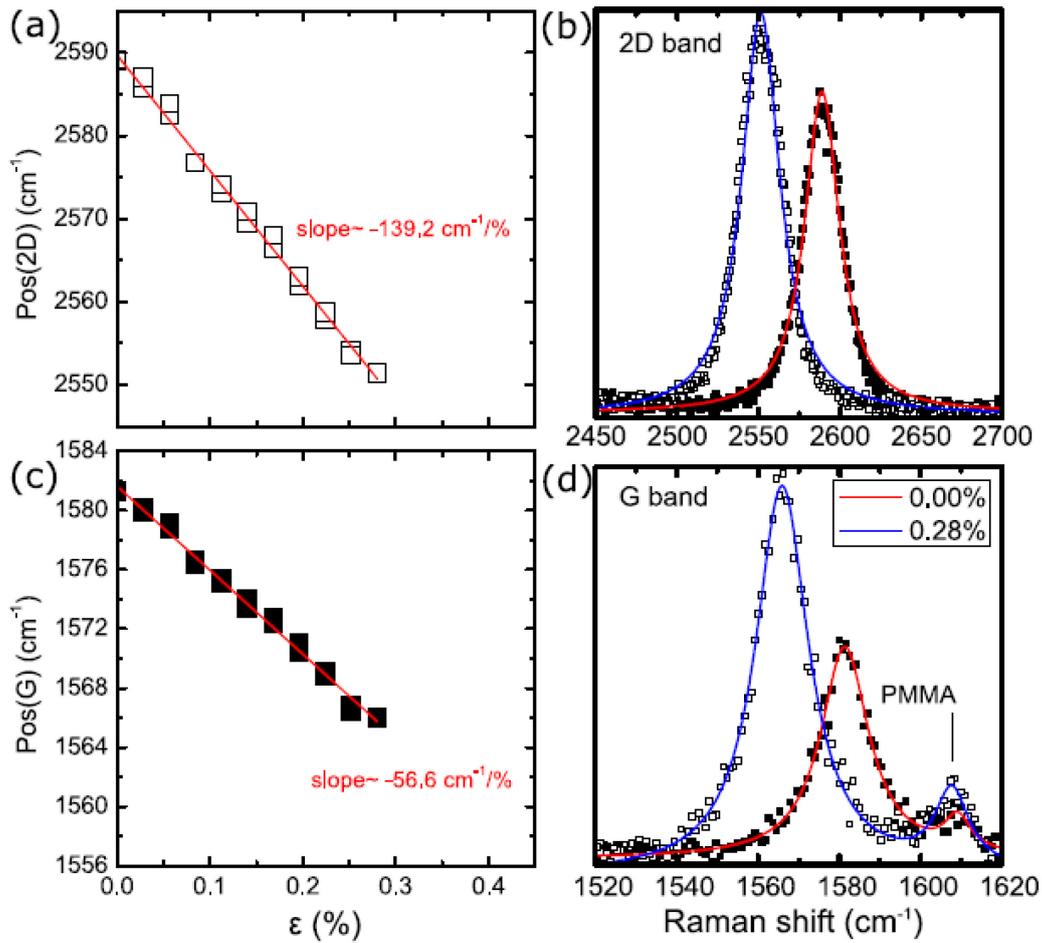

**Figure 2**. Position of the (a) 2D peak and (b) G peak versus applied strain. The insets show the dependence of the corresponding full-width-at-half-maximum on strain. Evolution of the (c) G and (d) 2D Raman spectra of graphene. Results are for monolayer graphene simply supported on PMMA/SU-8 substrate.

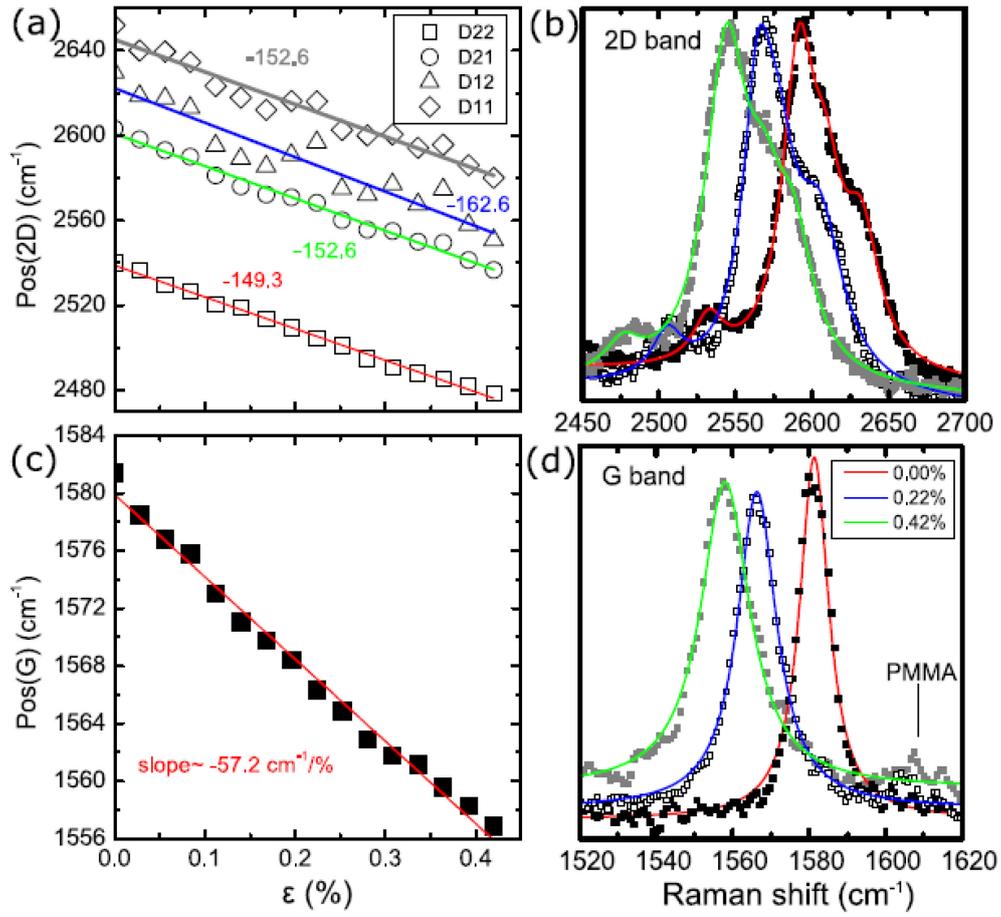

**Figure 3**. Position of the (a) four components of the 2D peak and (b) G peak versus applied strain. The inset shows the dependence of the G peak full width at half maximum on strain. Evolution of the (c) G and (d) 2D Raman spectra for bilayer graphene fully embedded in polymer.

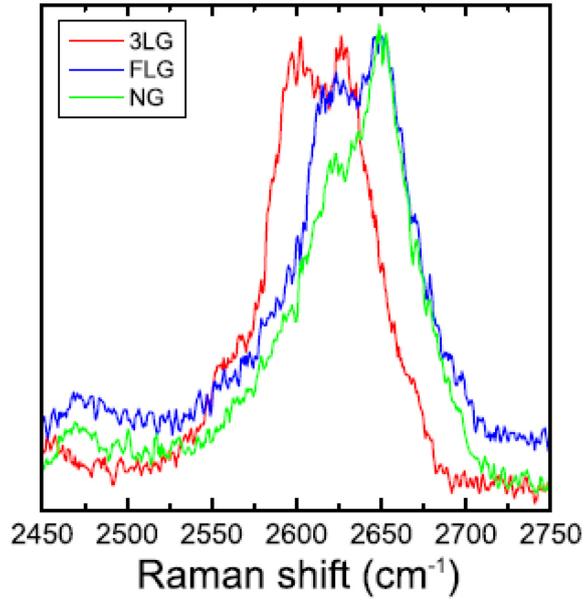

**Figure 4.** Representative 2D Raman spectra of the thicker graphene flakes, trilayer (3LG), few layer (FLG) and nanographite (NG)

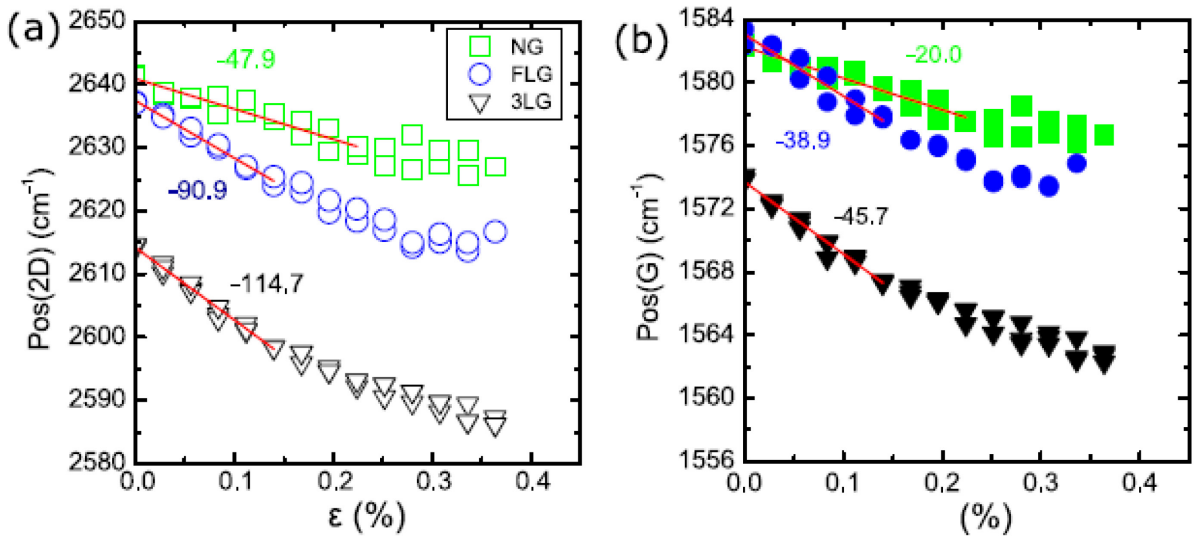

**Figure 5.** Position of the 2D peaks and G peaks versus applied strainfor (a), (b), trilayer graphene, (c), (d), few-layer graphene, and (e), (f) nanographite, respectively.

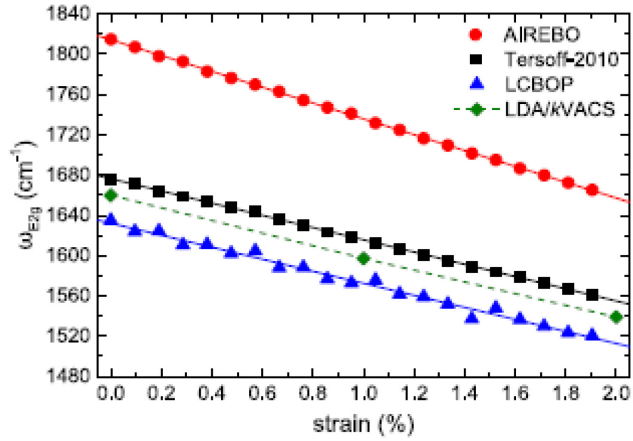

**Figure 6.** Evolution of the $E_{2g}$ phonon frequency upon equibiaxial strain calculated using the AIREBO (red circles), Tersoff-2010 (black squares), and LCBOP (blue triangles) potentials. The solid green rhombi are data from our DFT calculation within the LDA. In all cases the frequencies were calculated using the *k*VACS method (see Methods section) on MD simulations at a temperature of $T = 300$ K.

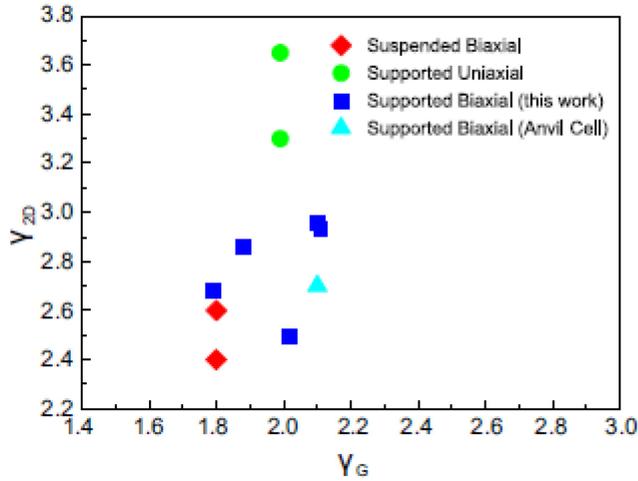

**Figure 7.** Experimental 2D vs G Grüneisen parameters, for various experimental setups.

**Table I.** Measured and calculated Grüneisen parameters and shift rates with applied strain of the G and 2D peaks.

| method | | biaxial strain or stress | $\gamma_G$ | $\Delta\omega_G/\varepsilon_{//}$ (cm$^{-1}$/%) | $\gamma_{2D}$ | $\Delta\omega_{2D}/\varepsilon_{//}$ (cm$^{-1}$/%) |
|---|---|---|---|---|---|---|
| uniaxial [4] | Supported | – | 1.99 | –63 | 3.55 | –191[§] |
| adhered on depression [16] | Supported | 0.066% | 2.4 | –77 | 3.8 | –203 |
| graphene bubble [15] | Suspended | ~1% | 1.8 | – | 2.6 | – |
| piezoelectric [17] | Supported | –0.15% to +0.1% | 1.80[§§] | –57.3[¶] | 2.98 | –160.3 |
| Blister [20] | Suspended | | 1.8 | –57 | 2.4 | –128 |
| Diamond anvil cell [21] | Supported | 3.5 GPa | 1.99[‖] | – | 2.7[‖] | – |
| Diamond anvil cell[22] | Supported | 6 GPa | 2.1[‡#] | – | 2.7 | – |
| monolayer, this work | Supported | 0.5% | 1.98±0.2 | –62.6±4 | 2.86±0.12 | –139 to –154 |
| DFT/LDA[23] | | 1% | 2.0 | –65[†] | – | – |
| DFT/LDA [4] | | 1% | 1.8 | –58 | 2.7 | –144 |
| DFT/GGA [24] | | –16% to +20% | 1.86 | –59 | – | – |
| DFT/GGA [25] | | 15% | 1.86[*] | –58.4[**] | – | – |
| DFT/GGA [26] | | 3% | – | –60 | – | –135 |
| DFT/LDA, this work[♀] | | 2.0% | 1.82 | –60.5 | – | – |
| AIREBO, this work[♀] | | 2.0% | 2.17 | –78.4 | – | – |
| Tersoff-2010, this work[♀] | | 2.0% | 1.80 | –60.7 | – | – |
| LCBOP, this work[♀] | | 2.0% | 1.91 | –59.7 | – | – |
| bilayer balloon [15] | | ~1.2%[§] | – | –56.6[¥¥] (34 cm$^{-1}$/bar) | – | –121.6 to –131.6[¥¥] (73–79 cm$^{-1}$/bar) |
| bilayer, this work | | 0.42% | 1.82 | –57.2 | – | –149 to –163 |
| trilayer, this work | | 0.34% | 1.45 | –45.7 | | –114.7 |
| few-layer, this work | | 0.34% | 1.23 | –38.9 | | –90.9 |
| nanographite, this work | | 0.34% | 0.62 | –19.9 | | –47.9 |

[¶] Strain range estimated using an estimated Grüneisen parameter.
[¥¥] Converted using the correspondence between the reported maximum differential pressure of 2 bar and the reported maximum achieved strain of 1.2%.
[§] Estimated.
[§§] Fixed; theoretical value taken from Ref 4.
[‖] The authors report agreement with the experimental and theoretical values from Ref. 4, for the G-peak and 2D peak respectively.
[‡] Calculated using an effective 3D bulk modulus of $B_{eff}$ = 600 GPa.
[*] Extracted from linear fit of data up to 3% strain.
[**] Calculated using a theoretical zero-strain $E_{2g}$ frequency at 1570.9 cm$^{-1}$.
[†] Calculated using a theoretical zero-strain $E_{2g}$ frequency at 1624 cm$^{-1}$.
[#] Calculated using a zero-strain $E_{2g}$ frequency of $\omega_G^0 = 1582$ cm$^{-1}$.
[♀] At finite temperature of $T$ = 300 K.

**Table II.** Experimental 2D and G Raman slopes for monolayer graphene.

| Substrate | 2D (cm$^{-1}$/%) | G (cm$^{-1}$/%) |
|:---:|:---:|:---:|
| PC | –141.0 | – |
| PC/SU-8 | –154.2 | – |
| PMMA | –153.8 | –66.35 |
| PMMA/SU-8 | –139.2 | –56.64 |
| PMMA/SU-8 | –152.3 | –66.55 |
| PMMA/SU-8 | –148.6 | –59.4 |
| **Average** | **–148.2±6** | **–62.3±5** |
| **Grüneisen** | $\gamma_{2D}$~ **2.86±0.12** | $\gamma_G$ ~ **1.97±0.15** |

**Table III.** Experimental average 2D and G peak Raman slopes for graphenes of various thicknesses.

| Number of layers | 2D (cm$^{-1}$/%) | G (cm$^{-1}$/%) |
|:---:|:---:|:---:|
| bilayer | –154.3 | –57.2 |
| trilayer | –114.7 | –45.7 |
| few-layer | –90.9 | –38.9 |
| nanographite | –47.9 | –19.9 |